\documentclass[letterpaper, 10 pt, conference]{ieeeconf}  

\usepackage{graphicx}
\usepackage{amssymb}
\usepackage{amsmath}
\usepackage{textcomp}
\usepackage{booktabs}
\usepackage{hyperref}
\usepackage{algorithmic}

\usepackage{todonotes}
\usepackage{mathtools}

\IEEEoverridecommandlockouts                              

\title{\LARGE \bf
Modality Bank: Learn multi-modality images across data centers without sharing medical data
}

\author{Qi Chang$^{1}$, Hui Qu$^{1}$, Zhennan Yan$^{2}$, Yunhe Gao$^{1}$, Lohendran Baskaran$^{3}$ and Dimitris Metaxas$^{1}$
\thanks{$^{1}$Qi Chang, Hui Qu, Yunhe Gao and Dimitris Metaxas are with Department of Computer Science, Rutgers University, Piscataway, New Jersey, USA
        {\tt\small qc58@rutgers.edu}}%
\thanks{$^{2}$Zhennan Yan is with the SenseBrain Research, Princeton, New Jersey, USA}%
\thanks{$^{3}$Lohendran Baskaran is with the Department of Cardiovascular Medicine, National Heart Centre Singapore, and Duke-National University Of Singapore}%
}

\begin{document}

\maketitle
\thispagestyle{empty}
\pagestyle{empty}

\begin{abstract}

Multi-modality images have been widely used and provide comprehensive information for medical image analysis. However, acquiring all modalities among all institutes is costly and often impossible in clinical settings. To leverage more comprehensive multi-modality information, we propose privacy secured decentralized multi-modality adaptive learning architecture named $\textit{ModalityBank}$. Our method could learn a set of effective domain-specific modulation parameters plugged into a common domain-agnostic network. We demonstrate by switching different sets of configurations, the generator could output high-quality images for a specific modality. Our method could also complete the missing modalities across all data centers, thus could be used for modality completion purposes.  The downstream task trained from the synthesized multi-modality samples could achieve higher performance than learning from one real data center and achieve close-to-real performance compare with all real images. 
\end{abstract}

\section{INTRODUCTION}

It is widely known that a sufficient amount of data plays a pivotal role in training a deep learning model~\cite{domingos2012few}. 
However, we still face big hurdles in terms of medical data sharing and collaboration for several reasons. 
Apart from the fact that the privacy policies such as HIPAA~\cite{annas2003hipaa,centers2003hipaa} and GDPR~\cite{regulation2018general,goddard2017eu} restrict the sharing of the patients' sensitive data, the heterogeneous nature of medical images by itself makes it more difficult to collaborate and analyze. 
Due to the different clinical acquisition protocols \cite{brown2018using,ellingson2015consensus} or various practical reasons across hospitals and countries, gathering all modalities among all institutes is a nontrivial task and sometimes even impossible.
As a result, such discrepancies hinder the machine learning model from learning the across-modality images~\cite{chen2019octopusnet,cheng2018deep} and ultimately hurt the performance. 

Multi-modality images, including MR imaging with several acquisition parameters, non-contrast/ contrast CT\cite{Ledezma2009,denecke2005comparison}, Ultrasound~\cite{horvat2018multimodality}, and PET(positron emission tomography)\cite{pichler2008multimodal}, can help to extract features from different perspectives and provide comprehensive information in medical image analysis. Nowadays many studies about medical cross-modality translation focus on image-to-image modality adaptation~\cite{yang2020mri,dou2018unsupervised}, and aiming at improving one single task performance like segmentation~\cite{chen2019synergistic,han2021deep} or classification\cite{lee2020cross}. However, it is still challenging to collect data and train all pairs of modality translations.
Therefore, a generative model that can adaptively generate multi-modality images for various downstream tasks' is worth exploring.

A good way of leveraging the private sensitive images is Federated Learning~\cite{kenecny2016federated,brisimi2018federated,bonawitz2019towards,li2019privacy}. The federated learning brings code to the patient data owners and only shares intermediate model training updates among them. However, different sites may have misaligned image modalities, which raises a realistic challenge for Federated Learning. Some recent methods~\cite{chang2020synthetic,qu2020learn} adopt a Decentralized Generative Adversarial Network(AsynDGAN) to address both sensitive data and continuous learning challenges. This architecture trains a central generator from the distributed discriminators across private data centers. The well-trained generator can act as an image provider to synthesize images for downstream tasks like segmentation or classification.
Though the AsynDGAN could learn to generate several image modalities~\cite{chang2020multi}, the number of modalities is limited by presetting the number of output channels of the generator.



Inspired by the above methods, we propose a new method on multi-modality adaptive learning with a privacy-secured solution, \textit{ModalityBank}. The \textit{ModalityBank} can not only generate multi-modality images for some downstream tasks but also be extended easily for more modalities.
Briefly, our contributions lie in three folds: 1) Proposed a privacy-secured decentralized multi-modality adaptive learning architecture, \textit{ModalityBank}. It learns a common domain-agnostic network and a set of effective domain-specific modulation parameters. We demonstrate by switching different sets of configurations, the generator could output high-quality images for a specific modality. 
2) \textit{ModalityBank} can synthesize realistic multi-modality images and complete the missing modalities across data centers.
3) The downstream task trained from the synthesized multi-modality samples could achieve higher performance than learning from real data of a single center and achieve close-to-ideal performance compared with using all real images from all sites.

\section{Methods}

\begin{figure*}[ht]
    \begin{center}
    \includegraphics[width=0.95\linewidth]{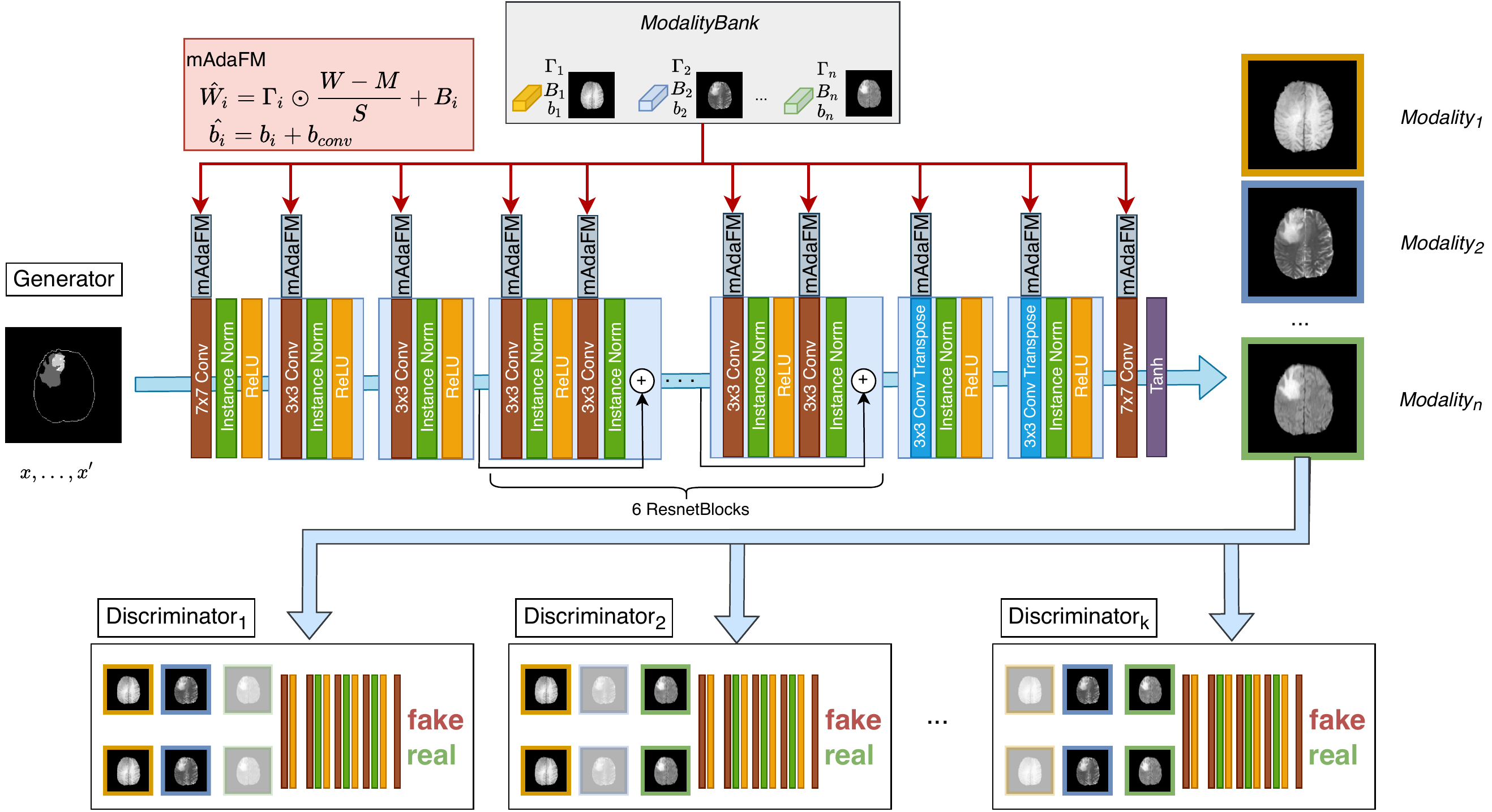}
    \end{center}
    \caption{The overall structure of ModalityBank. It contains three parts: sets of parameters $\Gamma_i$,$B_i$,$b_i$ for modality $i$,  a central generator $G$, multiple distributed discriminators $D_1, D_2, \cdots, D_k$ in each medical entity. 
    $G$ takes a task-specific input (segmentation masks in our experiments) and outputs multi-modality synthetic images by different adaptive parameters. 
    Each discriminator learns to differentiate between the real images of current medical entity and synthetic images from $G$. 
    The well-trained $G$ is then used as an image provider to train a task-specific model (segmentation in our experiments).
    }
\label{arch1}
\end{figure*}

Our proposed \textit{ModalityBank} is comprised of one domain-specific modulation parameters bank, one central generator and multiple distributed discriminators located in the different medical centers. In the following, we present the parameters bank, and then network architecture. 

\subsection{Domain-specific modulation parameters bank}
\label{modalitybank}

The generation of multi-modality data can be formulated as a style modulation task. Inspired from the style-transfer literature which alter the statistics of the features by firstly normalizing it with it's own mean and variance then performing affine transformation using the style image's mean and variance \cite{huang2017arbitrary}, we propose to use a modified adaptive filter modulation (mAdaFM)\cite{cong2020gan} to modulate the statistics of the weight in convolutional kernels to synthesis multi-modality images, even with severe style difference. 

To be precise, we introduce the reparameterizations of domain-specific modulation parameters bank as
$\Phi=\{\mathrm{\Gamma}_1,\mathrm{B}_1,\mathrm{b}_1, \mathrm{\Gamma}_2,\mathrm{B}_2,\mathrm{b}_2,\cdots, \mathrm{\Gamma}_n,\mathrm{B}_n,\mathrm{b}_n\}
$, $1\cdots n$ indicates $n$ types of modality images. The original convolutional layers:

\begin{equation}
    \mathrm{y}=f_{conv}(x;\mathrm{W},\mathrm{b}_{conv})
\end{equation}

could be present as:

\begin{equation}
\begin{multlined}
    \mathrm{y_i}=f_{conv}(x;\hat{\mathrm{W}}_i,\hat{\mathrm{b}}_i), i\in[1,n] \\
    \hat{\mathrm{W}}_i = \mathrm{\Gamma}_i\odot \frac{\mathrm{W}-\mathrm{M}}{\mathrm{S}}+\mathrm{B}_i, \quad \hat{\mathrm{b}}_i=\mathrm{b_i}+\mathrm{b}_{conv}
\end{multlined}
\end{equation}



where $\mathrm{M}, \mathrm{S}\in\mathbb{R}^{C_{out}\times C_{in}}$ denote the mean and standard deviation of the weight in the convolutional kernel, respectively. The $\mathrm{\Gamma}, \mathrm{B}\in\mathbb{R}^{C_{out}\times C_{in}}$ and $\mathrm{b}_{Conv}\in\mathbb{R}^{C_{out}}$ represent learnable modality-specific parameters. The generator was first pre-trained on one initial modality and then the $\mathrm{W}$ and $\mathrm{b}$ of each convolutional/deconvolutional layer are fixed. The learnable parameters $\Gamma_i, B_i, b_i$ are style parameters assigned to each modality $i$, and is trained to transform the fixed convolutional kernel to $\hat{\mathrm{W}},\hat{b}$ to synthesis images for the target modality.

The process of multi-modality synthesis is shown in Fig. \ref{arch1}, the pretrained generator is freezed, and the  learnable style parameters ($\gamma, \beta, b_{conv}$) in mAdaFM are used to modulate the Conv/Deconv layers. Therefore, we only need to store one generator along with a small set of style paramters for the synthesis of multi-modality images.

\subsection{Network architecture}

Our proposed \textit{ModalityBank} is comprised of one domain-specific modulation parameters bank described in section \ref{modalitybank}, one central generator and multiple distributed discriminators located in different medical entities.
An overview of the proposed architecture is shown in Figure~\ref{arch1}.
The central generator, denoted as $Generator$, takes task-specific inputs (e.g. segmentation masks in our use case) and generates synthetic images to fool the discriminators. 
Let $k$ denote the number of medical data centers that are involved in the learning framework. 
Our architecture ensures that $Discriminator_k$ deployed in the $k$-th medical entity only has the access to its local dataset, while not sharing any real image data outside the entity. During the learning process, Only synthetic images, masks, and losses are transferred between the central generator and the discriminators. Such design naturally complies with privacy regularization and keeps the patients' sensitive data safe. 

After training, the generator can be used as an image provider to generate training samples for some down-stream tasks. Assuming the distribution of synthetic images is same or similar to that of the real images, we can generate one unified large dataset which approximately equals to the union of all the datasets in medical entities. In this way, all private image data from each entity are utilized without sharing. In order to evaluate the synthetic images, we use the generated samples in segmentation tasks to illustrate the effectiveness of proposed \textit{ModalityBank}.

\section{Experiments}
In this section, we apply ModalityBank on a real MRI dataset, BraTS18, and evaluate it in two different settings. In the first setting, we use three heterogeneous data centers with three different modalities. Secondly, we explore the ModalityBank's adaptability to complete the missing modality across data centers. With different settings among the data centers and modalities, we could evaluate the performance of \textit{ModalityBank} towards a real-world scenario. We pre-trained the network based on the images from the different dataset(BraTS18 LGG). 
Without loss of generality, we adopt image segmentation as the down-stream task described in this paper.

\subsection{Dataset}


The BraTS2018 dataset comes from the Multimodal Brain Tumor Segmentation Challenge 2018~\cite{menze2014multimodal,bakas2017advancing}. 
All images are acquired from the three different sources: (1) The Center for Biomedical Image Computing and Analytics (CBICA) (2) The Cancer Imaging Archive (TCIA) data center (3) Data from other sites (Other). 
Each case has four types of MRI scan modalities (T1, T1c, T2 and FLAIR) and three types of tumor sub-region labels. All modalities have been aligned to a common space and resampled to 1mm isotropic resolution~\cite{bakas2018identifying}.
The 210 HGG cases in the challenge training set are split into train (170 cases) and test (40 cases) sets in our study since we have no access to the test data.

In our experiments, we evaluate our method to learn the distribution of all HGG cases across different data centers. In the GAN synthesis phase, all three labels are utilized to generate fake images. For segmentation, we focus on the whole tumor region (union of all three labels). The image dataset used in each experiment share one or multiple modalities.
Without loss of generality, we picked T1+T2+FLAIR, T1c+T2+FLAIR modalities respectively for the following two experiments.

\subsection{Experiment on multi-modal datasets}

\begin{figure*}[ht]
	\begin{center}
	    \includegraphics[width=0.13\linewidth]{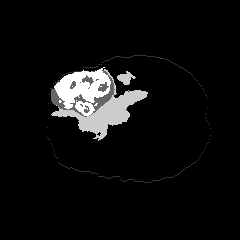}
	    \includegraphics[width=0.13\linewidth]{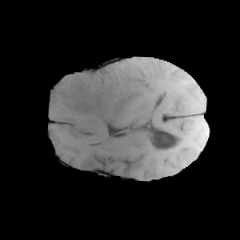}
	    \includegraphics[width=0.13\linewidth]{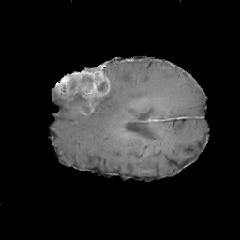}
	    \includegraphics[width=0.13\linewidth]{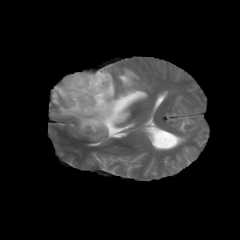}
	    \includegraphics[width=0.13\linewidth]{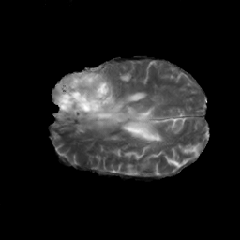}
	    \includegraphics[width=0.13\linewidth]{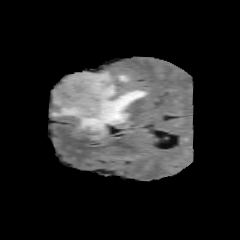}
	    \includegraphics[width=0.13\linewidth]{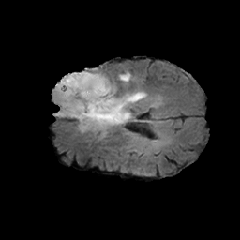}\\ 
	    \includegraphics[width=0.13\linewidth]{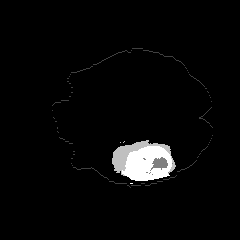}
		\includegraphics[width=0.13\linewidth]{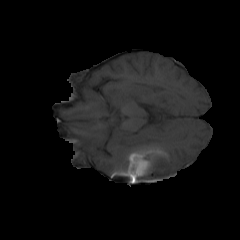}
		\includegraphics[width=0.13\linewidth]{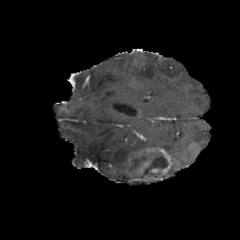}
		\includegraphics[width=0.13\linewidth]{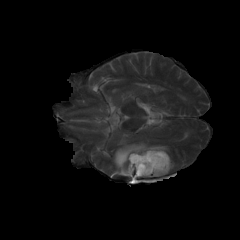}
		\includegraphics[width=0.13\linewidth]{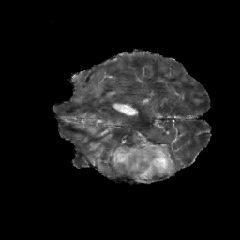}
		\includegraphics[width=0.13\linewidth]{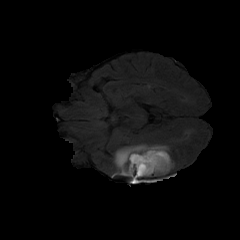}
	    \includegraphics[width=0.13\linewidth]{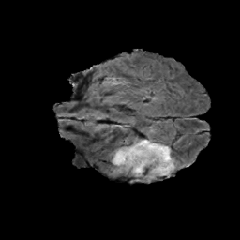}\\ 
		\begin{minipage}{0.13\linewidth}
	    \includegraphics[width=\linewidth,height=2.30cm]{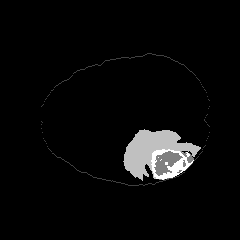}\\(a)Input
	    \end{minipage}
	    \begin{minipage}{0.13\linewidth}
	    \includegraphics[width=\linewidth,height=2.31cm]{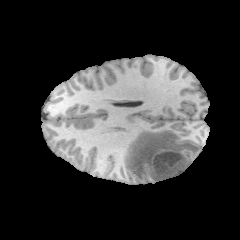}\\(e) Real T1
	    \end{minipage}
	    \begin{minipage}{0.13\linewidth}
	    \includegraphics[width=\linewidth,height=2.3cm]{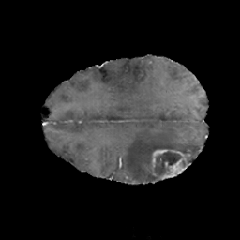}\\(b) Syn T1
	    \end{minipage}
	    \begin{minipage}{0.13\linewidth}
	    \includegraphics[width=\linewidth,height=2.31cm]{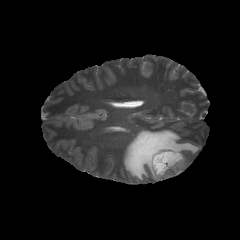}\\(f) Real T2
	    \end{minipage}
	    \begin{minipage}{0.13\linewidth}
	    \includegraphics[width=\linewidth,height=2.3cm]{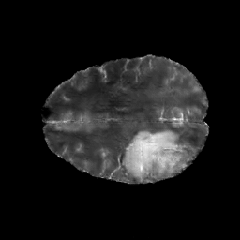}\\(c) Syn T2
	    \end{minipage}
	    \begin{minipage}{0.13\linewidth}
	    \includegraphics[width=\linewidth,height=2.3cm]{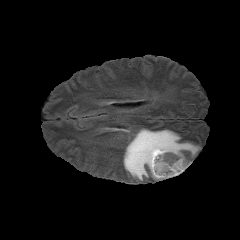}\\(g)RealFlair
	    \end{minipage}
	    \begin{minipage}{0.13\linewidth}
	    \includegraphics[width=\linewidth,height=2.3cm]{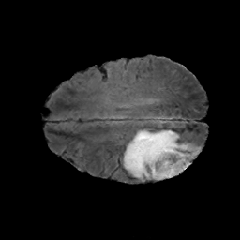}\\(d)SynFlair
	    \end{minipage}
	\end{center}
	\caption{The examples of multi-modality synthetic brain tumor images from the  \textit{ModalityBank}. (a) The input of the  \textit{ModalityBank}. (b)-(d) Synthetic multi-modality images of  \textit{ModalityBank}. (e)-(g) Real multi-modality images.} 
	\label{fig:syn:hgg-multimod}
\end{figure*}

\begin{table*}
    \caption{\label{tab:hgg-3}Brain tumor segmentation results over three heterogeneous and multi-modal (T1+T2+Flair) subsets.}
	\begin{center}
		\begin{tabular}{p{0.1\textwidth}p{0.1\textwidth}p{0.1\textwidth}p{0.1\textwidth}p{0.1\textwidth}}
			\toprule
			Method & Dice(\%) $\uparrow$ & Sens(\%) $\uparrow$  & Spec(\%)$\uparrow$  & HD95$\downarrow$ \\
			\midrule
			Real-All & 87.9\textpm8.5 & 85.6\textpm13.5 & 99.8\textpm0.3 & 10.51\textpm5.93 \\ 
			FedML-All & 87.3\textpm8.4 &	85.22\textpm14.9 &	99.8\textpm0.2 & 12.6\textpm0.2 \\ 
			\midrule
			Real-CBICA & 78.9\textpm19.6 & 75.7\textpm23.1 & 99.7\textpm0.2 & 16.45\textpm9.89 \\
			Real-TCIA & 77.2\textpm12.1 & 82.1\textpm16.1 & 99.3\textpm0.4 & 12.68\textpm4.95 \\
			Real-Other & 80.4\textpm12.9 & 80.7\textpm19.4 & 99.5\textpm0.3 & 23.33\textpm14.0 \\ 
			\midrule
			 AsynDGAN  & 82.0\textpm17.6 & 81.9\textpm22.0 & 99.5\textpm0.6 & 13.93\textpm10.0 \\
			\textbf{ModalityBank}  & 85.2\textpm10.9 & 82.4\textpm17.1 & 99.7\textpm0.2 & 14.66\textpm9.92 \\\midrule
			
		\end{tabular}
	\end{center}
\end{table*}
In this experiment, we show that our \textit{ModalityBank} can learn the distributions and generate realistic multi-modality medical images across heterogeneous data centers.
Specifically, the generator can generate realistic three channels (T1, T2, Flair) multi-modality images by learning from three heterogeneous data sources.

The training data is split into 3 subsets based on the different sources of the data described in \cite{menze2014multimodal}:
(1) Real-CBICA, 88 cases collected from CBICA.
(2) Real-TCIA, 102 cases collected from TCIA.
(3) Real-Other, 20 cases collected not from CBICA nor TCIA.


The brain tumor segmentation results on the test set are shown in Table~\ref{tab:hgg-3}.
The model trained using all real images (Real-All) is the ideal case scenario that we can access all data. It is our baseline and achieves the best performance. Compared with the ideal baseline, the performance of the models trained only using data in each medical entity (Real-CBICA, Real-TCIA, Real-Other) degrades a lot. 
We use the FedML.ai library~\cite{chaoyanghe2020fedml} for FedML-All experiment to train the segmentation model. It can make use of real images from all three subsets thus its performance is lightly lower than Real-All. 

AsynDGAN and our $\textit{ModalityBank}$ all produce synthetic images to train the segmentation model but differ in the way of training the GAN. 
Our $\textit{ModalityBank}$ and AsynDGAN both train the GAN in a distributed setting that is close to the real-world scenario. $\textit{ModalityBank}$ outperforms AsynDGAN because the domain-specific modulation parameters bank can better handle different modalities.
Our method can learn the information of all subsets during training, although the generator doesn't "see" the real images. Therefore, it outperforms all models learn using a single subset. 
Some examples of synthetic images from our method and corresponding real images are shown in Fig.~\ref{fig:syn:hgg-multimod}.
Worth noticing that the number of one modality configuration parameters is 2.5M while the number of all frozen source parameters is 21M. With smaller trainable parameters, the $\textit{ModalityBank}$ could learn and store the modalities configuration more efficiently. 

\subsection{Experiment on missing-modality datasets}

\begin{table*}
    \caption{\label{tab:hgg-4}Brain tumor segmentation results over three datasets with missing modality (T1c/T2/Flair)}
	\begin{center}
		\begin{tabular}
		{lcccc}
			\toprule
			Method & Dice(\%)~$\uparrow$ & Sens(\%)~$\uparrow$  & Spec(\%)~$\uparrow$  & HD95~$\downarrow$ \\
			\midrule
			Real-CBICA(n/a:T2) & 78.0\textpm23.4 & 74.5\textpm25.9 & 99.7\textpm0.2 & 15.47\textpm14.2 \\
			Real-TCIA(n/a:Flair) & 76.7\textpm15.3 & 72.8\textpm20.8 & 99.5\textpm0.8 & 15.64\textpm8.75 \\
			Real-Other(n/a:T1c) & 80.9\textpm14.1 & 79.3\textpm18.8 & 99.6\textpm0.2 & 16.74\textpm9.41\\ \midrule
			FedML-All & 82.9\textpm8.7 & 90.2\textpm13.1 & 99.2\textpm0.7 & 21.88\textpm11.52 \\
			\textbf{ModalityBank} & 85.8\textpm10.9 & 83.8\textpm16.6 & 99.7\textpm0.2 & 14.71\textpm7.99 \\
			\midrule
			\textbf{Completed-CBICA (syn:T2)}  & 83.0\textpm14.6 & 79.9\textpm19.0 & 99.7\textpm0.2 & 15.64\textpm9.93\\
			\textbf{Completed-TCIA (syn:Flair)}& 85.5\textpm10.4 & 83.3\textpm14.3 & 99.7\textpm0.1 & 15.02\textpm8.35  \\
            \textbf{Completed-Other (syn:T1c)} & 80.9\textpm15.3 & 80.6\textpm19.1 & 99.6\textpm0.2 & 16.93\textpm11.8 \\
			\bottomrule
		\end{tabular}
	\end{center}
\end{table*}

\begin{figure*}[ht]
	\begin{center}
		\includegraphics[width=\linewidth]{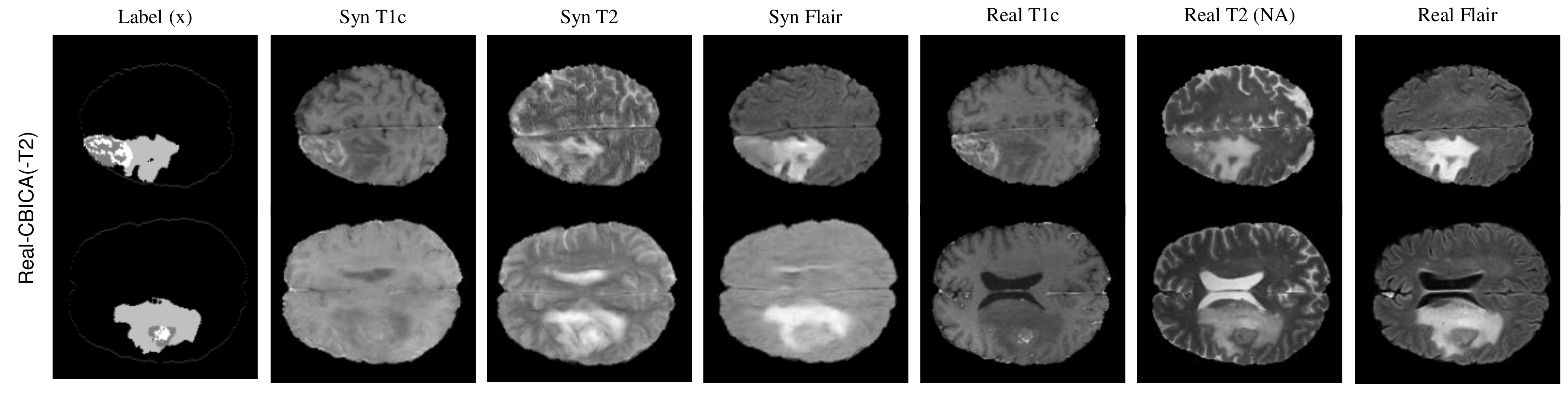}
		\includegraphics[width=\linewidth]{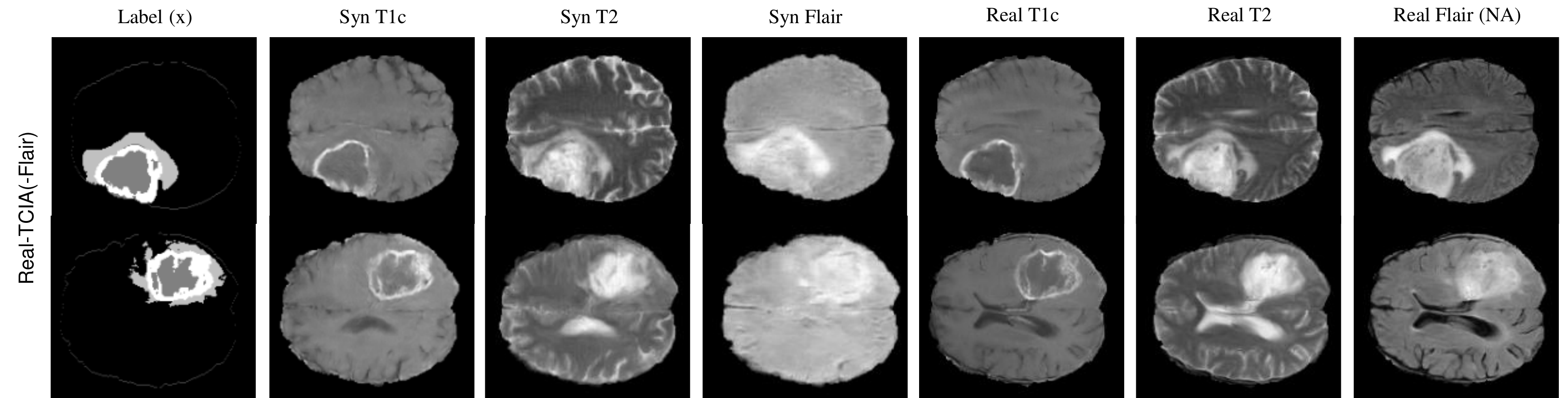}
		\includegraphics[width=\linewidth]{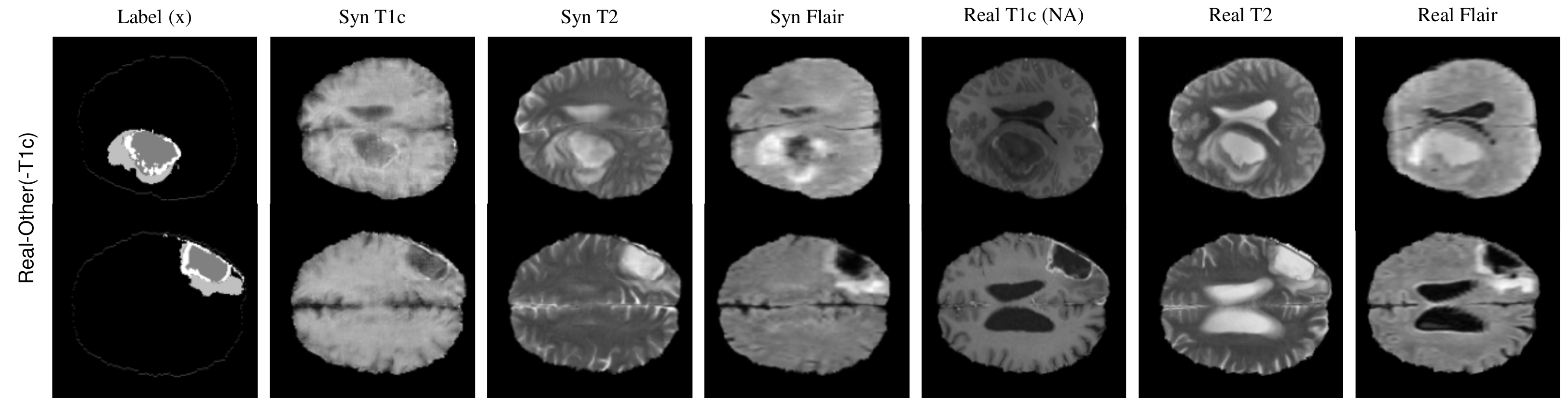}
	\end{center}
	\caption{The examples of synthetic brain tumor images by the  \textit{ModalityBank} after learning from multiple missing-modality datasets. `NA' column indicates the missing modality during training.}
	\label{fig:syn:missing}
\end{figure*}

In this section, we show that our \textit{ModalityBank} can learn the misaligned modality distribution and generate the complete multi-modality images. Specifically, the generator could generate realistic three channel(T1c, T2, Flair) multi-modality images while the real datasets don't provide one of the three modalities. 

The training data is split into 3 subsets based on the different sources of the data described in \cite{menze2014multimodal} and skip one of the modality described below:(1) Real-CBICA(n/a:T2) skip T2 modality. (2) Real-TCIA(n/a:Flair) skip Flair modality (3) Real-Other(n/a:T1c) skip T1c modality.


The brain tumor segmentation results about modality completion are shown in Table \ref{tab:hgg-4}. Overall, segmentation performance dropped when the segmentation network only learns from the real data center with missing modality while completed multi-modality images generated by \textit{ModalityBank}
could help the segmentation network to achieve much higher results. 
Our method has the best performance compared with the real images from one data center and the federated learning method. Though FedML-All could learn the real distribution across all data centers, the architecture couldn't adapt to the discrepancy of missing modalities, therefore it has the worse performance.

By providing the missing modality images for each of the datasets, the completed dataset would also outperform the counterpart of the real dataset. We also notice that T2 and Flair may contribute more to the whole tumor segmentation task since learning from the smallest subset Real-Other(n/a:T1c) achieves higher performance compared with learning from the other subsets with missing T2 or Flair. As a result, there is no significant difference between Completed-Other (syn:T1c) and Real-Other(n/a:T1c) by introducing the synthetic T1c images. 

We show some examples of synthetic images and corresponding real images in Fig.~\ref{fig:syn:missing}. In this figure, the 3 sections are corresponding to three data centers, respectively. The column of the real image labeled as NA (not available) indicates the missing modality in that center during the training of  \textit{ModalityBank}. The first observation is that our method can still learn to generate multiple modalities when centers have missing modality. We also notice that the synthetic images may not have the same global context as the real images, for example, the generated brains may have different shapes of ventricles. This is due to the lack of information about other tissues outside the tumor region in the input of the $G$.
On one hand, this variation is good for privacy preservation. On the other hand, for missing modality completion, the synthetic modality may have a different context from the real modalities.
However, this limitation seems not critical to our segmentation task, since the results in Table~\ref{tab:hgg-4} show clear improvement after the missing modality completion.

\subsection{Ablation Study}

In our ablation study, our network is pretrained from different datasets: a. BraTS18 LGG (lower grade glioma) dataset with T1c modality. b. BraTS18 HGG (high grade glioma) dataset with T1c modality.
c.M\&Ms (Multi-Centre, Multi-Vendor \& Multi-Disease) Cardiac Image Segmentation Challenge dataset. 

\begin{table}
    \caption{\label{tab:ablation}Ablation study for different pre-trained model}
    \setlength\tabcolsep{4pt}
	\begin{center}
		\begin{tabular}{lcccc}
			\toprule
			Method & Dice(\%)~$\uparrow$ & Sens(\%)~$\uparrow$  & Spec(\%)~$\uparrow$  & HD95~$\downarrow$ \\
			\midrule
			Real-CBICA & 78.0\textpm23.4 & 74.5\textpm25.9 & 99.7\textpm0.2 & 15.47\textpm14.2 \\
			Real-TCIA & 77.2\textpm12.1 & 82.1\textpm16.1 & 99.3\textpm0.4 & 12.68\textpm4.95 \\
			Real-Other & 80.4\textpm12.9 & 80.7\textpm19.4 & 99.5\textpm0.3 & 23.33\textpm14.0 \\ 
			\midrule
			\multicolumn{5}{l}{\textbf{ModalityBank Pre-trained on}}\\
			\midrule
			BraTS18 LGG & 85.2\textpm10.9 & 82.4\textpm17.1 & 99.7\textpm0.2 & 14.66\textpm9.92 \\
			BraTS18 HGG & 84.4\textpm14.9 & 81.2\textpm17.9 & 99.8\textpm0.2 & 13.95\textpm13.07 \\
			M\&Ms & 84.0\textpm12.0 & 80.7\textpm17.0 & 99.7\textpm0.2 & 18.32\textpm12.4 \\
			
			\midrule
		\end{tabular}
	\end{center}
\end{table}

The results clearly show the \textit{ModalityBank} can achieve a better performance with flexible types of pre-trained model. The pre-trained model extract useful features rather than the prior knowledge which can be used in the target domain.  

\section{Conclusion and future work}
\label{sec:conclude}

In this work, we proposed a privacy secured decentralized multi-modality adaptive learning architecture named $\textit{ModalityBank}$. By applying multiple domain-specific modulation parameters, our method demonstrated improving multi-modality image quality and higher performance of the downstream task. In addition, we showed that $\textit{ModalityBank}$ is an efficient way to complete missing modalities and thus unifies the medical images from different data centers with various modalities. 
It's also worth to mention that we choose the different sets of modalities in two experiments to demonstrate the generality of the proposed method. In future, we will extend to support different number of channels and 3D images volumes.

\bibliographystyle{IEEEtran}

\bibliography{root}

\begin{thebibliography}{10}
\providecommand{\url}[1]{#1}
\csname url@rmstyle\endcsname
\providecommand{\newblock}{\relax}
\providecommand{\bibinfo}[2]{#2}
\providecommand\BIBentrySTDinterwordspacing{\spaceskip=0pt\relax}
\providecommand\BIBentryALTinterwordstretchfactor{4}
\providecommand\BIBentryALTinterwordspacing{\spaceskip=\fontdimen2\font plus
\BIBentryALTinterwordstretchfactor\fontdimen3\font minus
  \fontdimen4\font\relax}
\providecommand\BIBforeignlanguage[2]{{%
\expandafter\ifx\csname l@#1\endcsname\relax
\typeout{** WARNING: IEEEtran.bst: No hyphenation pattern has been}%
\typeout{** loaded for the language `#1'. Using the pattern for}%
\typeout{** the default language instead.}%
\else
\language=\csname l@#1\endcsname
\fi
#2}}

\bibitem{domingos2012few}
P.~M. Domingos, ``A few useful things to know about machine learning.''
  \emph{Commun. acm}, vol.~55, no.~10, pp. 78--87, 2012.

\bibitem{annas2003hipaa}
G.~J. Annas \emph{et~al.}, ``Hipaa regulations-a new era of medical-record
  privacy?'' \emph{New England Journal of Medicine}, vol. 348, no.~15, pp.
  1486--1490, 2003.

\bibitem{centers2003hipaa}
C.~for Disease~Control, Prevention, \emph{et~al.}, ``Hipaa privacy rule and
  public health. guidance from cdc and the us department of health and human
  services,'' \emph{MMWR: Morbidity and mortality weekly report}, vol.~52, no.
  Suppl. 1, pp. 1--17, 2003.

\bibitem{regulation2018general}
P.~Regulation, ``General data protection regulation,'' \emph{Intouch}, 2018.

\bibitem{goddard2017eu}
M.~Goddard, ``The eu general data protection regulation (gdpr): European
  regulation that has a global impact,'' \emph{International Journal of Market
  Research}, vol.~59, no.~6, pp. 703--705, 2017.

\bibitem{brown2018using}
A.~D. Brown and T.~R. Marotta, ``Using machine learning for sequence-level
  automated {MRI} protocol selection in neuroradiology,'' \emph{Journal of the
  American Medical Informatics Association}, vol.~25, no.~5, pp. 568--571,
  2018.

\bibitem{ellingson2015consensus}
B.~M. Ellingson, M.~Bendszus, J.~Boxerman, D.~Barboriak, B.~J. Erickson,
  M.~Smits, S.~J. Nelson, E.~Gerstner, B.~Alexander, G.~Goldmacher,
  \emph{et~al.}, ``Consensus recommendations for a standardized brain tumor
  imaging protocol in clinical trials,'' \emph{Neuro-oncology}, vol.~17, no.~9,
  pp. 1188--1198, 2015.

\bibitem{chen2019octopusnet}
Y.~Chen, J.~Chen, D.~Wei, Y.~Li, and Y.~Zheng, ``Octopusnet: A deep learning
  segmentation network for multi-modal medical images,'' in \emph{International
  Workshop on Multiscale Multimodal Medical Imaging}.\hskip 1em plus 0.5em
  minus 0.4em\relax Springer, 2019, pp. 17--25.

\bibitem{cheng2018deep}
X.~Cheng, L.~Zhang, and Y.~Zheng, ``Deep similarity learning for multimodal
  medical images,'' \emph{Computer Methods in Biomechanics and Biomedical
  Engineering: Imaging \& Visualization}, vol.~6, no.~3, pp. 248--252, 2018.

\bibitem{Ledezma2009}
\BIBentryALTinterwordspacing
C.~J. Ledezma and M.~Wintermark, ``\BIBforeignlanguage{eng}{Multimodal ct in
  stroke imaging: new concepts},'' \emph{\BIBforeignlanguage{eng}{Radiologic
  clinics of North America}}, vol.~47, no.~1, pp. 109--116, Jan 2009. [Online].
  Available: \url{https://pubmed.ncbi.nlm.nih.gov/19195537}
\BIBentrySTDinterwordspacing

\bibitem{denecke2005comparison}
T.~Denecke, B.~Rau, K.-T. Hoffmann, B.~Hildebrandt, J.~Ruf, M.~Gutberlet,
  M.~H{\"u}nerbein, R.~Felix, P.~Wust, and H.~Amthauer, ``Comparison of ct, mri
  and fdg-pet in response prediction of patients with locally advanced rectal
  cancer after multimodal preoperative therapy: is there a benefit in using
  functional imaging?'' \emph{European radiology}, vol.~15, no.~8, pp.
  1658--1666, 2005.

\bibitem{horvat2018multimodality}
N.~Horvat, M.~S. Rocha, A.~L. Chagas, B.~C. Oliveira, M.~P. Pacheco, M.~A.
  Binotto, N.~M. Ikari, D.~C. Paranagu{\'a}-Vezozzo, H.~M. Leao-Filho, J.~R.~T.
  Vicentini, \emph{et~al.}, ``Multimodality screening of hepatic nodules in
  patients with congenital heart disease after fontan procedure: role of
  ultrasound, arfi elastography, ct, and mri,'' \emph{American Journal of
  Roentgenology}, vol. 211, no.~6, pp. 1212--1220, 2018.

\bibitem{pichler2008multimodal}
B.~J. Pichler, M.~S. Judenhofer, and C.~Pfannenberg, ``Multimodal imaging
  approaches: Pet/ct and pet/mri,'' \emph{Molecular Imaging I}, pp. 109--132,
  2008.

\bibitem{yang2020mri}
Q.~Yang, N.~Li, Z.~Zhao, X.~Fan, I.~Eric, C.~Chang, and Y.~Xu, ``Mri
  cross-modality image-to-image translation,'' \emph{Scientific reports},
  vol.~10, no.~1, pp. 1--18, 2020.

\bibitem{dou2018unsupervised}
Q.~Dou, C.~Ouyang, C.~Chen, H.~Chen, and P.-A. Heng, ``Unsupervised
  cross-modality domain adaptation of convnets for biomedical image
  segmentations with adversarial loss,'' \emph{arXiv preprint
  arXiv:1804.10916}, 2018.

\bibitem{chen2019synergistic}
C.~Chen, Q.~Dou, H.~Chen, J.~Qin, and P.-A. Heng, ``Synergistic image and
  feature adaptation: Towards cross-modality domain adaptation for medical
  image segmentation,'' in \emph{Proceedings of the AAAI Conference on
  Artificial Intelligence}, vol.~33, no.~01, 2019, pp. 865--872.

\bibitem{han2021deep}
X.~Han, L.~Qi, Q.~Yu, Z.~Zhou, Y.~Zheng, Y.~Shi, and Y.~Gao, ``Deep symmetric
  adaptation network for cross-modality medical image segmentation,''
  \emph{arXiv preprint arXiv:2101.06853}, 2021.

\bibitem{lee2020cross}
J.~Lee and R.~M. Nishikawa, ``Cross-organ, cross-modality transfer learning:
  Feasibility study for segmentation and classification,'' \emph{IEEE Access},
  vol.~8, pp. 210\,194--210\,205, 2020.

\bibitem{kenecny2016federated}
J.~Konečný, H.~B. McMahan, F.~X. Yu, P.~Richtárik, A.~T. Suresh, and
  D.~Bacon, ``{Federated Learning: Strategies for Improving Communication
  Efficiency},'' 2016.

\bibitem{brisimi2018federated}
T.~S. Brisimi, R.~Chen, T.~Mela, A.~Olshevsky, I.~C. Paschalidis, and W.~Shi,
  ``Federated learning of predictive models from federated electronic health
  records,'' \emph{International journal of medical informatics}, vol. 112, pp.
  59--67, 2018.

\bibitem{bonawitz2019towards}
K.~Bonawitz, H.~Eichner, W.~Grieskamp, D.~Huba, A.~Ingerman, V.~Ivanov,
  C.~Kiddon, J.~Kone{\v{c}}n{\`y}, S.~Mazzocchi, H.~B. McMahan, \emph{et~al.},
  ``Towards federated learning at scale: System design,'' \emph{arXiv preprint
  arXiv:1902.01046}, 2019.

\bibitem{li2019privacy}
W.~Li, F.~Milletar{\`\i}, D.~Xu, N.~Rieke, J.~Hancox, W.~Zhu, M.~Baust,
  Y.~Cheng, S.~Ourselin, M.~J. Cardoso, \emph{et~al.}, ``Privacy-preserving
  federated brain tumour segmentation,'' in \emph{International Workshop on
  Machine Learning in Medical Imaging}.\hskip 1em plus 0.5em minus 0.4em\relax
  Springer, 2019, pp. 133--141.

\bibitem{chang2020synthetic}
Q.~Chang, H.~Qu, Y.~Zhang, M.~Sabuncu, C.~Chen, T.~Zhang, and D.~N. Metaxas,
  ``Synthetic learning: Learn from distributed asynchronized discriminator
  {GAN} without sharing medical image data,'' in \emph{Proceedings of the
  IEEE/CVF Conference on Computer Vision and Pattern Recognition}, 2020, pp.
  13\,856--13\,866.

\bibitem{qu2020learn}
H.~Qu, Y.~Zhang, Q.~Chang, Z.~Yan, C.~Chen, and D.~Metaxas, ``Learn distributed
  gan with temporary discriminators,'' in \emph{European Conference on Computer
  Vision}.\hskip 1em plus 0.5em minus 0.4em\relax Springer, 2020, pp. 175--192.

\bibitem{chang2020multi}
Q.~Chang, Z.~Yan, L.~Baskaran, H.~Qu, Y.~Zhang, T.~Zhang, S.~Zhang, and D.~N.
  Metaxas, ``Multi-modal asyndgan: Learn from distributed medical image data
  without sharing private information,'' \emph{arXiv preprint
  arXiv:2012.08604}, 2020.

\bibitem{huang2017arbitrary}
X.~Huang and S.~Belongie, ``Arbitrary style transfer in real-time with adaptive
  instance normalization,'' in \emph{Proceedings of the IEEE International
  Conference on Computer Vision}, 2017, pp. 1501--1510.

\bibitem{cong2020gan}
Y.~Cong, M.~Zhao, J.~Li, S.~Wang, and L.~Carin, ``Gan memory with no
  forgetting,'' \emph{arXiv preprint arXiv:2006.07543}, 2020.

\bibitem{menze2014multimodal}
B.~H. Menze, A.~Jakab, S.~Bauer, J.~Kalpathy-Cramer, K.~Farahani, J.~Kirby,
  Y.~Burren, N.~Porz, J.~Slotboom, R.~Wiest, \emph{et~al.}, ``The multimodal
  brain tumor image segmentation benchmark ({BRATS}),'' \emph{IEEE transactions
  on medical imaging}, vol.~34, no.~10, pp. 1993--2024, 2014.

\bibitem{bakas2017advancing}
S.~Bakas, H.~Akbari, A.~Sotiras, M.~Bilello, M.~Rozycki, J.~S. Kirby, J.~B.
  Freymann, K.~Farahani, and C.~Davatzikos, ``Advancing the cancer genome atlas
  glioma {MRI} collections with expert segmentation labels and radiomic
  features,'' \emph{Scientific data}, vol.~4, p. 170117, 2017.

\bibitem{bakas2018identifying}
S.~Bakas, M.~Reyes, A.~Jakab, S.~Bauer, M.~Rempfler, A.~Crimi, R.~T. Shinohara,
  C.~Berger, S.~M. Ha, M.~Rozycki, \emph{et~al.}, ``Identifying the best
  machine learning algorithms for brain tumor segmentation, progression
  assessment, and overall survival prediction in the brats challenge,''
  \emph{arXiv preprint arXiv:1811.02629}, 2018.

\bibitem{chaoyanghe2020fedml}
C.~He, S.~Li, J.~So, M.~Zhang, H.~Wang, X.~Wang, P.~Vepakomma, A.~Singh,
  H.~Qiu, L.~Shen, P.~Zhao, Y.~Kang, Y.~Liu, R.~Raskar, Q.~Yang, M.~Annavaram,
  and S.~Avestimehr, ``Fedml: A research library and benchmark for federated
  machine learning,'' \emph{arXiv preprint arXiv:2007.13518}, 2020.

\end{thebibliography}

\end{document}